\documentclass[acmsmall,authorversion,nonacm]{acmart}   

\AtBeginDocument{%
  \providecommand\BibTeX{{%
    \normalfont B\kern-0.5em{\scshape i\kern-0.25em b}\kern-0.8em\TeX}}}

\usepackage{url}
\usepackage{doi}

\usepackage{currfile}
\usepackage{textcomp}

\usepackage{enumitem}

\begin{document}


\title{From Trade-only to Zero-Value NFTs: The Asset Proxy NFT Paradigm in Web3}

\author{Denis Avrilionis}
\affiliation{%
\institution{Compellio SA}
\city{Luxembourg}
\country{Luxembourg. }
\href{mailto:denis@compell.io}{denis@compell.io}
}

\author{Thomas Hardjono}
\affiliation{%
\institution{MIT Connection Science \& Engineering}
\city{Cambridge, MA}
\country{USA. }
\href{mailto:hardjono@mit.edu}{hardjono@mit.edu}
}

\begin{abstract}
Many implementations of smart contracts available in NFT marketplaces today 
allow the modification of NFT token attributes, without any specific mechanism to control the consistency with off-chain metadata. 
We believe this is a weakness in overall design of NFTs today.
We propose a computation model called the {\em Asset Proxy NFT} 
that guarantees the
consistency between the NFT token (on-chain) and 
its corresponding asset metadata (off-chain). 
In general, the proposed model can be applied to any type of NFT 
that requires immutability or controlled mutability of metadata.
A second contribution of this paper
is the notion of the {\em NFT design patterns}
which recognizes that a coherent framework for dealing with hybrid assets is required,
and that for specific hybrid-asset deployments,
suitable technological components must be utilized under the framework.
\\
~~\\
{\bf \today}

~~
\end{abstract}





\maketitle


\section{Introduction: Hybrid Assets}

There is today a growing interest in the use of the Non-Fungible Tokens (NFT)
on the blockchain to represent assets
that are {\em hybrid assets}.
These are asset that generally consist of two parts,
namely a real-world valuable goods/resources (off-chain)
and an NFT that represents the asset on a blockchain.
The notion here is that the ownership of the NFT on-chain 
would legally imply (i.e. effect)
the corresponding ownership of the asset off-chain.
This interest in hybrid assets dovetails into the current broad discourse
regarding the future {\em Web3} as the ``Internet of Value''
-- versus the current Web2 as the ``Internet of Communications''~\cite{Wood2018-web3}.
However,
there are today several challenges facing blockchain technology
in its current iteration as a part of the Web3 proposition~\cite{Das-NFT-Security-2021,OReilly2021-web3,Moxie2022-web3}.

We believe that many of these these issues arise largely from an incorrect view of assets.
An {\em asset-centric view} of the world must take into account the fact
that in the majority of cases the assets and their associated economic value 
originate from outside the blockchain.
Just as the design of the TCP/IP Internet was centered 
around communications survivability~\cite{Clark88},
we believe Web3 must be centered around assets and not around the mechanics of blockchain technology.

Although Web3 is currently going through its own hype-cycle~\cite{SmithCearley2022-Gartner-Web3},
Web2 computing infrastructures are not going away anytime soon.
For the past few decades 
many modern financial institutions have invested heavily in IT infrastructures 
-- including cloud computing --
to enable the digitization of the various financial services (B2B and B2C).
Thus, another oft-overlooked aspect of the Web3 discourse is the need
for the future Web3 infrastructures to integrate securely with
these existing Web2 systems and networks. 
An asset-centric view must allow for the bridging between the Web2 world and the future Web3.

In this paper
we propose an {\em asset proxy paradigm} for NFTs
where the focus is placed on the asset itself
and where the NFT construct is seen as a proxy (on-chain) for the asset (off-chain).
This paradigm in itself introduces several new challenges,
notably the need for a bilateral synchronization mechanism to be utilised
that ensures the perpetual consistency of the state between the NFT on-chain and the asset off-chain.
One key implication is that
any metadata that supports the utilization and operations of the NFT
must be protected against unauthorized modifications.

A second contribution of this paper
is the notion of the {\em NFT design patterns}
which recognizes that (i) a coherent framework for dealing with hybrid assets is required,
and that
(ii) for specific hybrid-asset deployments,
suitable technological components must be utilized under the framework.

\section{NFT - state of the practice}

 The current state of the practice puts the emphasis on smart contract programming related to the on-chain management of the lifecycle of NFT tokens (e.g. as in the case of ERC721~\cite{EntrikenShirley2018} 
 via dedicated methods defined in the specification, for minting, transferring, or burning tokens). Although such aspects are very important, the topics related to the synchronized and consistent management of the off-chain asset linked to the NFT token are of equal importance, especially in the context of physical assets (manufactured products, luxury items, etc.).

Below we present an example of an existing NFT implementation to illustrate some key aspects of the technology, before introducing our asset proxy NFT paradigm in the following sections.

\subsection{A detailed example}
\label{sec:NFT-state}

To illustrate our approach, we will use the example the Christie's \href{https://www.christies.com/lot/lot-6345173}{\em Beeple NFT}\footnote{\href{https://www.christies.com/lot/lot-6345173}{https://www.christies.com/lot/lot-6345173}} that was sold in November 2021 for USD\$28.9 Million. We borrow this example from~\cite{miller2021}.

By navigating the Ethereum mainnet we can see the token's \href{https://etherscan.io/address/0xa4c38796C35Dca618FE22a4e77F4210D0b0350d6}{\em smart contract address}\footnote{\href{https://etherscan.io/address/0xa4c38796C35Dca618FE22a4e77F4210D0b0350d6}{https://etherscan.io/address/0xa4c38796C35Dca618FE22a4e77F4210D0b0350d6}}. The Beeple token is the token number {\em One} in the smart contract. The token has a unique {\bf {\em attribute}} (we use the term ``attribute'' for all {\em on-chain} information related to an NFT token). The NFT attribute (the \verb|tokenURI| according to ERC721) is the URL link:
\begin{center}
\href{https://metadata.human-one.xyz/1}{\em https://metadata.human-one.xyz/1}
\end{center}
That link points to the following JSON {\bf {\em metadata}} file stored in IPFS (we use the term ``metadata'' for all {\em off-chain} information related to a NFT token):
\begin{verbatim}
{
 "image":"https://nft.human-one.xyz/Ukraine_22b55e18faae73ad86ce32cd.png",
 "animation_url":"https://nft.human-one.xyz/Ukraine_22b55e18faae73ad86ce32cd.mp4",
 "external_url":"https://human-one.xyz",
 "description":"millions of voices suddenly cried in terror and were suddenly silenced.",
 "name":"HUMAN ONE",
 "background_color":"000000",
 "days_journeyed":150,
 "location":"broken future",
 "attributes":[{"trait_type":"Location","value":"broken future"}]
}
\end{verbatim}

\begin{figure}[h]
\centering
\includegraphics[width=0.7\textwidth, trim={0.0cm 0.0cm 0.0cm 0.0cm}, clip]{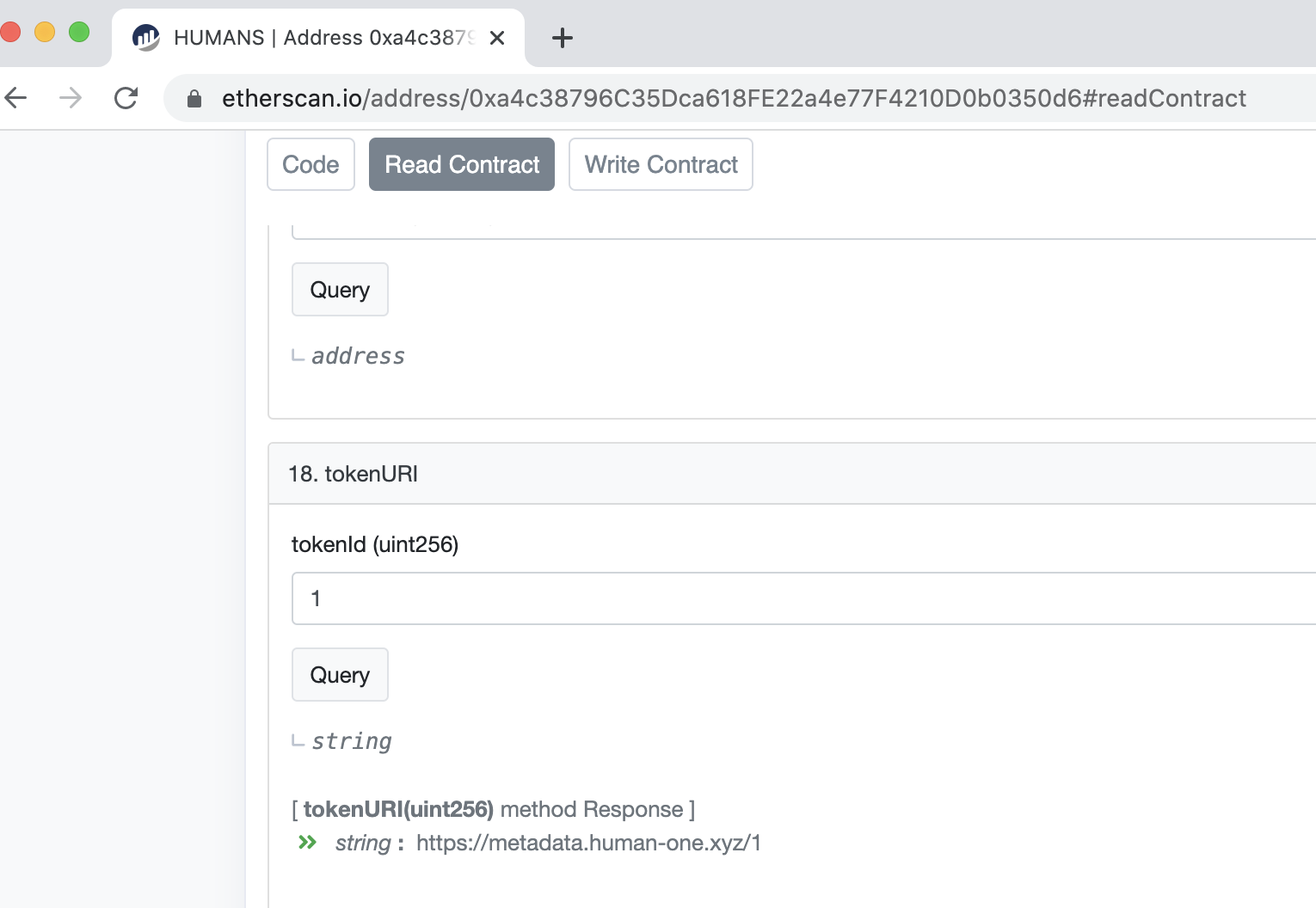}
\caption{Example of the Beeple NFT with attribute information (ERC-721)}
\label{fig:figure1label}
\end{figure}

Except for some very specific cases where both the NFT token and the related asset are both stored on-chain (e.g. see~\cite{arthaus2022}), NFTs typically have an off-chain part that is (or points to) the asset related to the specific NFT token. As we can see in the example above, the actual asset related to the (on-chain) NFT number {\em One} is the (off-chain) mp4 animation file available at the location:
\begin{center}
\href{https://nft.human-one.xyz/Ukraine\_22b55e18faae73ad86ce32cd.mp4}{https://nft.human-one.xyz/Ukraine\_22b55e18faae73ad86ce32cd.mp4}
\end{center}

Many implementations of smart contracts available in NFT marketplaces today 
(including, for example, OpenSea) allow the modification of NFT token attributes, without any specific mechanism to control the consistency with off-chain metadata. 
We believe this is a weakness in overall design of NFTs today.
More specifically,
if the NFT token metadata 
-- especially the URLs or other off-chain identifiers 
pointing to the asset associated with the token (like the URL in the metadata pointing to the MP4 file in the above example) --
is unintentionally changed or maliciously replaced, 
then the NFT token would point to a wrong or non-existing asset. 
Obviously,
such inconsistencies in the on-chain/off-chain states
are not acceptable, especially in the context of more complex machine-to-machine interactions, beyond simple token transfers in marketplaces (for example in the case of asset state changes in manufacturing / production / supply chain processes). 

A major requirement is the perpetual consistency between
the on-chain token and the off-chain asset related to the NFT token.  
In other words,
independent of whether this off-chain part is stored on a dedicated server,
a privately-hosted IPFS installation,
another off-chain storage mechanism, 
or even a traditional web2 application, 
there is the strong requirement 
to maintain the full consistency between the on-chain token and the off-chain asset.      

In order to overcome the potential for such inconsistencies,
we propose a computation model called {\bf {\em ``Asset Proxy NFT''}} that guarantees
consistency between the NFT token (on-chain) and its corresponding metadata (off-chain). In general,
the proposed model can be applied to any type of NFT 
that requires immutability (or controlled mutability) of metadata.

\subsection{Vocabulary}
\label{sec:vocabulary}

In order to provide clarity in the use of technical terminology,
in the remainder of this work, we use some specific terms that are defined as follows: 
\begin{itemize}[topsep=4pt,itemsep=1pt, partopsep=4pt, parsep=4pt]

\item A NFT token is {\bf {\em identified}} by:
\begin{itemize}
\item the chain in which the smart contract is deployed;
\item the address of the smart contract that manages the NFT;
\item the unique ID of the NFT inside the smart contract.
\end{itemize}

\item The NFT token has {\bf {\em Attributes}} (we borrow this term from the widely
deployed RFC2459~\cite{rfc2459}) stored on-chain. These attributes are usually links to off-chain information related to the token (e.g. \verb|tokenURI| in the example above)

\item The NFT token has {\bf {\em Metadata}} stored off-chain (JSON data in the example above). These metadata are usually under the responsibility of the {\bf {\em Asset Provider}} who is the {\bf {\em party}} (physical person or legal entity) that is responsible / accountable / liable with respect to the asset related to NFT token. Metadata can be stored in any off-chain medium, either centralized or decentralized. Metadata contain, among other, information related to the asset related to the NFT token.    

\item The {\bf {\em Asset}} that is related to the NFT token (the MP4 file in the example above) is most often stored off-chain. The management of the asset itself is also under the responsibility of the asset provider. There might be cases of {\bf {\em Hybrid Assets}} i.e. assets that also exist in physical form (manufactured products, luxury items, etc.). In such case additional redirections between the online representation of the asset and the physical product might also exist (e.g. online serial number associated to the corresponding physical product item).   

\item The asset provider might be the {\bf {\em Creator}} of the asset or might act as {\bf {\em Asset Custodian}}, holding and managing the asset on behalf of the creator ({\em Christie's} in the example above might be seen as an asset custodian acting on behalf of {\em Beeple}, the creator of the MP4 video).

\item Through the transfer of the NFT, 
the creator can sell the asset to another party who becomes the {\bf {\em Owner}} of the asset.  

\item The link between on-chain and off-chain information is usually unilateral from on-chain to off-chain as in the example above and is implemented via hyper-links stored as attributes on-chain. We refer to this link as {\bf {\em Resolution Mechanism}} allowing to retrieve off-chain asset information from on-chain NFT information.

\item A more sophisticated {\bf {\em Bilateral Synchronization Mechanism}} might be necessary in more complex cases where changes in the off-chain state of the asset might affect the metadata and the attributes of the NFT. A typical example is when off-chain assets evolve over time, for example due to changes in terms of production process, supply chain, tax/custom clearance, etc. Such synchronization mechanism must cater for both {\em off-chain to on-chain} and {\em on-chain to off-chain} modifications, always maintain consistency between the two worlds.  

\end{itemize}

Finally, it is very important to note the following: {\em Throughout this paper when we refer to an {\bf {\em asset}} we actually mean the specific {\bf {\em instance}} of a unique object that could be identified in an unambiguous way from another object instance of the same class / kind / type}.

\subsection{Limitations of the current NFT Model for Hybrid Assets}

One crucial issue pertains to the information
that permit the connection between an NFT and the off-chain asset instance
to be established. An important issue is the availability (persistence)
and the integrity protection of the metadata
against unauthorized modification.
Such unauthorized changes to the metadata
may result in the situation where 
there is a mismatch in the {\em resolution mechanism} (e.g. NFT attribute pointing to the wrong metadata, or metadata containing erroneous information about the asset).
This problem is particularly acute for the case of hybrid assets.

We summarize these challenges as follows:
\begin{itemize}[topsep=4pt,itemsep=1pt, partopsep=4pt, parsep=4pt]

\item	{\em Prevention of simultaneous on-chain double-spending}:
Currently it is difficult (impossible)
for a potential buyer of a hybrid asset to obtain assurance that
there is a one-to-one correspondence between 
the asset's representation off-chain and on-chain.

In other words,
there is currently no way for the potential buyer to obtain assurance
that the same off-chain asset is not made into several NFTs
at different blockchain networks (or even the same blockchain).

\item	{\em Detection of unauthorized modification of metadata}:
There is today a lack of mechanism to prevent the unauthorized modification 
of the metadata that refers to the off-chain asset instance.
This points to the need for a mechanism to achieve 
a {\em controlled mutability} of asset-related metadata.

\item	{\em Standard mechanism for hybrid assets}:
There is currently a lack of a standardized mechanisms for hybrid assets
that permit a synchronized correlation between
state changes on-chain with state changes off-chain (and vice-versa).

Thus, for example,
chain-specific operations (e.g. burn token) that are meaningful to tokens on a blockchain
may not have an equivalent operations off-chain -- and currently
there are no mechanisms to report this fact.

\item	{\em Strong identification of Asset Providers}:
The concept of ``decentralized trust'' (or ``trustless-ness'')
may be applicable to consensus-making on a blockchain,
but the concept may not be applicable to functions/actions off-chain. 

Thus, an asset provider as legal entity must be identifiable,
auditable, and held accountable in the case of any litigation related to the asset.

\end{itemize}

\section{The Asset Proxy NFT Paradigm}

The primary goal of the {\bf {\em Asset Proxy NFT paradigm}} introduced in this paper is to provide the means to achieve
technical {\em state consistency} between off-chain assets and the NFTs as an on-chain representations of these assets. 

We believe that this strong correlation or binding between the off-chain world and on-chain
world is core for the viability of blockchain technology and NFTs
as the future decentralized means for economic activity among people.
This is notably a crucial requirement for hybrid assets.

This implies that the mechanical means used
to attain this perpetual {\em bilateral synchronization} must be separated
from the economic valuation mechanisms used to determine 
the supply/demand pricing\footnote{The topic of the high cost of transactions
due to speculative buying of tokens have been discussed in~\cite{HardjonoLipton2021-FinTech}}.
We strongly believe that maintaining consistency between off-chain assets 
and on-chain tokens will be a core requirement for the new generation Web3 infrastructures.

\subsection{Architecture}

Figure~\ref{fig:assetProxyArchitecture} below presents a high-level component view of the Asset Proxy NFT Paradigm. 
The {\em Smart Contract} managing the NFT token(s) implements a {\em Resolution} mechanism that maps the
NFT {\em Attributes} to the {\em Asset Metadata}. 
Such a mechanism could be as simple as a URL pointing to the off-chain 
{\em Metadata Repository} that holds the {\em Asset Metadata}.  
The Asset Metadata contain references to the {\em Asset} that is managed 
via an {\em Asset Custody} service. 
A {\em Bilateral Synchronization} component controlled by the Creator of the Asset 
ensures full consistency among the state of the Asset, 
its Metadata and the Attributes of the NFT token.

\begin{figure}[h]
\centering
\includegraphics[width=0.9\textwidth, trim={0.0cm 0.0cm 0.0cm 0.0cm},clip]{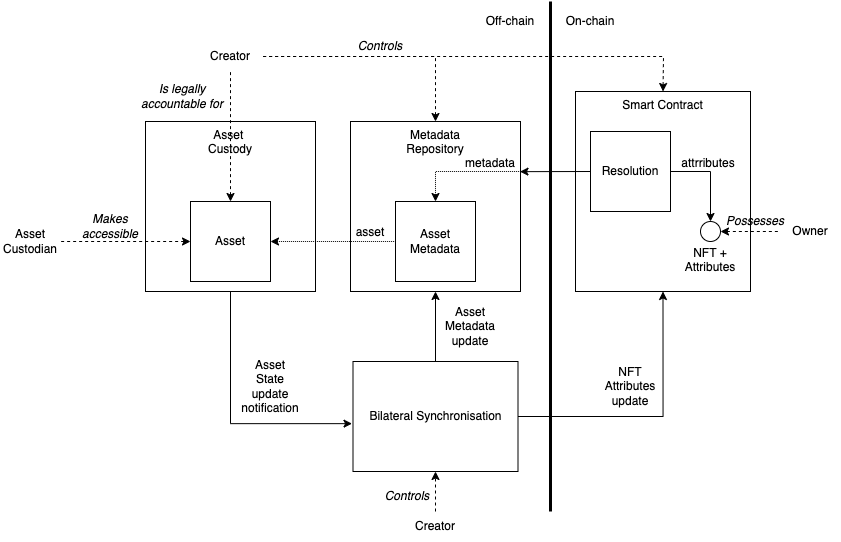}
\caption{Architecture of the Asset Proxy NFT Paradigm}
\label{fig:assetProxyArchitecture}
\end{figure}

\subsection{The Beeple NFT revisited}

If we view the Beeple NFT from the perspective of the Asset Proxy NFT Paradigm,
we see the following aspects:
 
\begin{itemize}[topsep=4pt,itemsep=1pt, partopsep=4pt, parsep=4pt]

\item The {\em Asset Custody} service is the end point that serves the MP4 file:\\ \href{https://nft.human-one.xyz/Ukraine\_22b55e18faae73ad86ce32cd.mp4}{https://nft.human-one.xyz/Ukraine\_22b55e18faae73ad86ce32cd.mp4}

\item The {\em Metadata Repository} is the end point that serves the JSON metadata:\\ \href{https://metadata.human-one.xyz/1}{https://metadata.human-one.xyz/1}. 

\item The {\em Resolution} mechanism is implemented in an implicit way via the \verb|tokenURI| field associated with the token (as defined in ERC721).

\item There are no {\em Bilateral Synchronization} mechanisms utilized since there are 
no modifications possible of the asset off-chain, 
and thus the on-chain information is not affected by any off-chain event.  
 
\end{itemize}

\subsection{Orchestration}

Below we describe three main sequences of interaction among the various components of the paradigm related to initialization, new asset deployment and update of the state of the asset (off-chain).

\medskip
\subsubsection{Services Initialization}~\smallskip

\begin{enumerate}[topsep=4pt,itemsep=1pt, partopsep=4pt, parsep=4pt]

\item The {\em Asset Custody} service is first
initialized by either the Creator itself or by a third-party custodian.

\item The {\em Metadata Repository} service is then initialized, 
with all configuration elements that are needed to communicate 
with the Asset Custody service (in case information about the asset is required). 

\item The {\em Smart Contract} is then deployed. 
As part of the initialization procedure the smart contract must configure 
the metadata {\em Resolution} mechanism (e.g. define the Metadata Repository service endpoint) in an immutable way.
(nb. the reconfiguration of the way metadata are related to token attributes 
should be avoided; if a reconfiguration is required or unavoidable 
then it must be performed in a secure and auditable manner by an authorized entity).

\end{enumerate}

\medskip
\subsubsection{New Asset deployment}~\smallskip

\begin{enumerate}[topsep=4pt,itemsep=1pt, partopsep=4pt, parsep=4pt]

\item The {\em Creator} places the asset under custody via the Asset Custody service. 
This might imply not only data updates in some storage mechanism but also legal due diligence from the custodian side, verification of intellectual property, additional notarization of ownership certificates,~etc. 

\item The {\em Metadata Repository} is updated with the asset metadata. 

\item The {\em Smart Contract} is updated, for example to mint the NFT token associated 
with the asset, set the attributes of the token to point to the correct metadata, etc.

\item The {\em Bilateral Synchronization} component is updated 
to correlate the asset identifier, the identifier of the asset metadata, as well as the identifier of the token.
  
\end{enumerate}

\medskip
\subsubsection{Off-chain update of asset state}~\smallskip

\begin{enumerate}[topsep=4pt,itemsep=1pt, partopsep=4pt, parsep=4pt]

\item The {\em Asset Custody} service notifies the {\em Bilateral Synchronization} component about the new state of the asset 

\item If required, the {\em Bilateral Synchronization} component updates the metadata to reflect the new state of the asset. 

\item If required, the {\em Bilateral Synchronization} component updates the attributes of the NFT. 

\end{enumerate}

\subsection{About identification of off-chain assets and on-chain tokens}

In this section we tackle a specific topic related to object identification that is of paramount importance for hybrid assets, since correlation of off-chain and on-chain identifiers in guaranteeing the overall consistency of the information.  
\smallskip
\subsubsection{Persistent Digital Identifiers for Physical Assets}~\\

With the rise of the ``Internet of Value'' where physical assets
can be represented digitally (e.g. as tokens on a blockchain)
and traded in the digital space (e.g. crypto-exchanges, metaverse, etc.)
there also arises the question of how to ensure that 
identifiers that are associated
with the physically assets can be represented uniquely
in the digital space.
This is relevant not only from a logistics point of view (i.e. supply-chain tracking),
but also, from a business survivability perspective (i.e. detecting counterfeits).
Identifiers associated with physical assets (e.g. product serial numbers)
can often exist for many decades,
even after the asset's creator ceases to exist (i.e. product manufacturer closes business).

In the Beeple example above there is a physical element associated with the NFT. However, as it is mentioned the special notice accompanying the NFT sales\footnote{\href{https://www.christies.com/lot/lot-6345173}{https://www.christies.com/lot/lot-6345173}} {\em ``The Artwork is neither stored nor embedded in the NFT, but is accessible through the NFT. 
The Physical Element displays a copy of the Artwork. [...] The NFT and Physical Element may be separated and owned by different persons. If you are the holder of the NFT or Physical Element, you may transfer the NFT or Physical Element to a third party, but, after you do so, your license to the Artwork will immediately terminate.''} Clearly, Beeple's decision was to completely dissociate the two elements. 
In the absence of clear identification of the physical asset 
the issue of maintaining perpetual and synchronized links between 
the NFT and the physical asset would have been impossible.  

In the digital space, the need for persisting a digital identifier goes beyond ensuring global uniqueness
of the byte-string identifier, but also ensuring that meaningful contextual metadata associated with the digital identifier are also available online.
In the approach presented in this paper we assume that a {\em Metadata Repository} (e.g. a service endpoint available on a particular location on the Internet) can map a digital identifier to a particular set of contextual information regarding the asset. Our approach thus shapes the notion of the {\em identifier architecture} and brings several important aspects with regards to identifiers that must persist in the digital space and in the real-world:
\begin{itemize}[topsep=4pt,itemsep=1pt, partopsep=4pt, parsep=4pt]

\item	{\em Control over identifier issuance}:
There needs to be an unambiguous entity that controls the mechanism
(i.e. protocol) to issue identifiers. For example, the ISBN numbering scheme~\cite{Wikipedia-ISBN} provides a means to associate a unique identifier to a published book

\item	{\em Control over metadata}:
In many cases the true value of an identifier lies in the metadata associated
with the identifier\cite{Hardjono2019-IEEECommsMagazine}.
Thus,
there must be mechanisms (technical or legal)
that ensures that the entity who issues an identifier
has control also over the issuance of the metadata. In the case of the ISBN numbering scheme the identifier is based on unique metadata
about the book (e.g., author, book title, etc.)

\item	{\em Control over resolution mechanisms}:
In the digital space,
there must be systems that guarantee the resolution of an identifier
to the associated metadata. 
This brings the question of the persistence of the metadata itself. In the case of the ISBN numbering scheme metadata are registered to a known authority (e.g., Library of Congress in the United States).

\item	{\em Survivable persistence of identifier and metadata}:
There must be a persistence scheme for metadata
that ensures that the metadata {\em survives} the issuer.
That is,
in some cases, an identifier and metadata in the digital space 
must be available for a very long term
(e.g. decades),
even beyond the lifetime of the issuing/controlling entity.

\end{itemize}

\smallskip
\subsubsection{On-chain Identifiers}~\\

Regarding on-chain identifiers, as already defined in the vocabulary section \ref{sec:vocabulary} above, the identification of NFTs is unique by construction. Typically, the combination of (i) the blockchain network identifier, (ii) the address of the smart contract managing the NFT and (iii) the unique ID of the token ensures global uniqueness of the identity of the NFT \cite{caip19}.  

In the Asset Proxy NFT paradigm, 
the NFT attributes are directly related to the metadata of the physical asset. 
By invoking a smart contract to make use of an NFT, 
the caller -- namely the Asset Provider -- is (legally) claiming 
that it is utilizing the identifier of the associated physical asset 
in the real-world (e.g. serial number of product). 
This could be seen as an {\em existential claim} of the identifier off-chain.

\smallskip
\subsubsection{Binding between Metadata and Attributes}~\\
\label{subsec:Binding}

In order to achieve off-chain / on-chain consistency in hybrid assets, specific mechanisms may be implemented:
\begin{itemize}[topsep=4pt,itemsep=1pt, partopsep=4pt, parsep=4pt]

\item	To enforce binding between the off-chain asset and the on-chain token, it is important that metadata records should contain both (i) an explicit reference to the on-chain identifier of the NFT and (ii) the off-chain identifier of the asset. 

\item Access to the {\em Metadata Repository} might be subject to authentication and authorization of the caller
(i.e. entity seeking access to the metadata record).

\item The {\em Bilateral Synchronization} mechanism must contain explicit (eventually immutable) reference to the smart contract that manages the NFTs associated to the off-chain asset. 

\end{itemize}

\section{Asset Proxy NFT Design Patterns: From Tradeable to Zero-Value NFTs}

The inclusion of asset-specific characteristics as NFT attributes 
within a smart contract and the permitted operations (as defined in the code of the smart contract) leads to the possibility of {\em design patterns}
to be created for certain types/classes of hybrid assets.

This permits an {\em asset-centric} view of the hybrid world to be developed, where the possible changes of state in the real-world asset
takes equal (or more) importance than what technical functions can be achieved using a blockchain. An asset-centric view also takes into account the legal and societal aspects of human economic activities, and the consequent ``assets'' they produce.

Thus, the pattern of behavior that is permitted to occur on an asset
is determined as part of the design of the associated NFT. The design of the NFT behavior takes into consideration several aspects of the asset lifecycle including non-technical matters like legal aspects or other policy concerns (e.g. taxation). Such permitted behavior then determines the choice of technological implementation of the NFT. This is in contrast to prevailing trend today, where a {\em blockchain-centric} view of the world determines the operations permissible on a token. For example, some token operations (e.g. ``burn'' token) has no real-world equivalent and may thus be unsuitable for hybrid assets. 
Today's blockchain-centric view amplified by the Oracle problem inherent to the design of blockchain technology \cite{AvrilionisHardjono2021Arxiv} is a major limitation in the design of NFTs for hybrid assets. We believe that NFT-related technology is only at its beginnings. The current trend that focuses on NFTs primarily as a means for trade is limiting both in terms of technical scope as well as in terms of business added-value. 

The Asset Proxy NFT paradigm is a step in the direction of a holistic architecture 
considering both off-chain and on-chain worlds. 
Given the broader technological scope of the paradigm 
there are several patterns of use of NFT technology that would be required 
to cover the needs of the hybrid asset space. 
We believe that a natural evolution of the current NFT technology landscape 
would be the creation of specific, well defined {\bf {\em NFT design patterns}}. 
Such patterns would specialize the Asset Proxy NFT paradigm 
in particular cases focusing on targeted but generic business or technical issues. 
Below we enumerate a list of NFT design patterns we have already identified. 
We expect this list to grow as the NFT usage becomes more widespread and 
NFT technology becomes more mainstream across industries.

The NFT design patterns presented below are described according to the four essential elements that define a design pattern \cite{GoF}, namely: 
\begin{itemize}[topsep=4pt,itemsep=1pt, partopsep=4pt, parsep=4pt]

\item the {\em pattern name}, which is a handle to describe the design problem, its solutions, and its consequences in a brief description.

\item the {\em problem}, which explains the context and describes when to apply the pattern.

\item the {\em solution}, which describes how the various elements of the Asset Proxy NFT paradigm are specialized in terms of their relationships, responsibilities, and collaborations.

\item the {\em consequences}, which are the results and trade-offs of applying the pattern.

\end{itemize}

\subsection{Trade-only NFT pattern}

\begin{itemize}[topsep=4pt,itemsep=1pt, partopsep=4pt, parsep=4pt]

\item {\bf {\em Trade-only NFT}}: Several marketplaces for trading unique assets 
(especially in digital form) are available today. 
When it comes to hybrid assets, the NFTs could be used to facilitate trade and 
thus increase the liquidity of the market. 
A Trade-only NFT implements the usual ``plain vanilla'' NFT that is common
today in most marketplaces.

\item {\bf {\em Problem}}: While buying an asset online is straightforward today, 
it is more difficult to implement the proper trading of assets 
as the ownership (as well as the authenticity) of the asset is difficult to establish in peer-to-peer mode. 
Moreover, the ownership must be clearly established before the asset gets delivered to the owner. 

\item {\bf {\em Solution}}: When an NFT associated with an off-chain asset
is to be created,
its metadata {\em must} bind (cryptographically) the unique identification 
of the off-chain asset (e.g. serial number) with the on-chain NFT identifier.
The ownership of the asset could then be instantly proven 
by the sole possession of the NFT token. 

When an owner decides to access the off-chain asset,
the token is burned on-chain and the off-chain asset is delivered (i.e. physically) to the owner. 
Burning of the token should be implemented as an irreversible action and implies the end of the trade. 
The custodian of the asset would then ship the off-chain asset to the owner who burned the token.

\item {\bf {\em Consequences}}: The Trade-only pattern has a number of
inherent limitations that is a consequences of the inherent nature of
the pattern.
More specifically,
the NFT has no notion or knowledge about about the current state of the asset (off-chain). 
As a consequence,
it is simply impossible for the owner to get updates regarding the state of the asset
only by observing the state of the NFT token on the blockchain. 

If, for example, the token is related to an asset to be produced in the future 
(e.g. a luxury item that will be produced after the token sales), 
then there is no way for the owner to be informed 
about the asset state through the attributes of the NFT token. 

In a sense, the NFT token limited to only cover
 "secondary market like" ownership transfer operations. 
In the case of off-chain assets, to avoid the double-spending problem,
the legitimate owner of the asset can obtain possession of the asset,
which implies the burning of the NFT.
The non-trading actions, such as the shipment of the asset to the owner (i.e. the last owner who burned the token) 
is a pure off-chain operation (which cannot be handled via smart contract programming).    

\end{itemize}

\subsection{Cross-Chain Transferable NFT pattern}

\begin{itemize}[topsep=4pt,itemsep=1pt, partopsep=4pt, parsep=4pt]

\item {\bf {\em Cross-Chain Transferable NFT}}: Some off-chain assets 
(e.g. assets with long lifecycles, such as high-value product components
or other mechanical/system engineering artefacts) 
may undergo changes of ownership in their long lifetime.
This means that the asset's association with its NFT
on a given blockchain may need to change.
For example,
when the asset changes ownership its NFT token may need
to be moved to another blockchain (e.g. the blockchain that is preferred/selected
by its new owner).

Thus, a core requirement in this pattern is the continuity and the consistency in the management of the state of the asset
across different blockchains.
This requirement is at the very heart of the
asset-centric view that we stated above.

\item {\bf {\em Problem}}: Today there are several industry-specific 
(vertically integrated)
blockchains created by market leaders. 
Some examples include Aura (Prada consortium)~\cite{Rascouet2021-PradaBlockchain} 
in the luxury industry, 
Quorum (ConsenSys) in the financial industry~\cite{Irrera2020}, 
Tradelens (Maersk/IBM)~\cite{Miller2019-Maersk} in supply chain management,
and others. 
Moreover, the rise of CBDCs and the increased interest among financial institutions 
in native-digital assets may drive interest in transferable NFTs,
which could be freely traded across blockchains operated/controlled
by groups of financial institutions. 

Thus,
a key problem is the need for assets in one blockchain
to be transferable to a different blockchain.
This may require the NFT in the origin blockchain
to be burned and an equivalent new NFT to be generated
on the destination blockchain~\cite{IETF-draft-hardjono-gateways}.

\item {\bf {\em Solution}}: Similar to the case of the {\em Trade-only NFT}, 
a bilateral association between the off-chain asset and the NFT must be established
in order that the ownership of the asset can be instantly proven 
by the possession of the NFT token. 
Whenever an asset transfer has to be performed 
the {\em Bilateral Synchronization} mechanism would burn the NFT 
on the origin blockchain and would create 
an equivalent new NFT on the destination blockchain. 
As part of the burning operation, the NFT on the origin blockchain would update its attributes to contain the reference to the identity of 
the newly create NFT on the destination blockchain~\cite{HardjonoLipton-IEEETEMS-2019}. 
The asset metadata should also be updated to maintain 
the history of the previous NFTs so both the off-chain and the on-chain information remain perpetually synchronized.

\item {\bf {\em Consequences}}: The {\em Bilateral Synchronization} mechanism 
responsible for the transfer -- namely the burning
of the NFT in the origin blockchain followed immediately 
by creation of the equivalent NFT in the destination blockchain --
must do so in an {\em atomic} manner.

In other words, 
either the new target NFT is created in the destination blockchain 
(and the source NFT in the origin blockchain in burned),
or no change occurs to the association between
the source NFT and the asset.
There cannot be an ``in between'' (intermediate) state where two NFTs
exists simultaneously in both the origin blockchain
and the destination blockchain.
The smart contract supporting cross-chain transferable tokens 
must prevent any operation on the NFT (e.g. change of ownership) during transfer. 
In case of failure in the sequence of operations the whole transfer should be reverted and the source NFT should remain unchanged.

\end{itemize}

\subsection{Hidden metadata NFT pattern}

\begin{itemize}[topsep=4pt,itemsep=1pt, partopsep=4pt, parsep=4pt]

\item {\bf {\em Hidden metadata NFT}}: There are use-cases or circumstances 
where the information about the asset must be protected, or even hidden. 
In these cases, both the on-chain information about the NFT
and the off-chain asset metadata must not reveal the asset information.
In order to access this information,
users must obtain access credentials from the creator of the asset.

\item {\bf {\em Problem}}: Some industries are utilizing NFTs for 
product authenticity (counterfeit detection) and supply-chain traceability. 
However,
in some cases the quantity of the product (i.e. number of units produced)
may be considered as sensitive information,
and therefore, must only be disclosed to authorized users or entities.

As another example, in some cases only confirmed bidders should be able
to retrieve metadata about the asset (e.g. in the case of a by-invitation only auction).
This is because unauthorized disclosure of such information
could be exploited by other bidders
(e.g. bidders could scan all assets,
and then bid only for assets with certain characteristics, for instance only for those with the highest rarity factor in case of assets built based on genetic algorithms).
Thus, in these cases the NFT attributes and the asset metadata 
must be disclosed selectively only to the approved entities.

\item {\bf {\em Solution}}: The attribute of the NFT should only carry
an opaque hash that uniquely identifies the serialized metadata 
related to the asset associated with the NFT. 
Correspondingly,
the asset metadata should contain a reverse reference to the NFT identifier. 
When a user with the appropriate credentials wishes 
to retrieve the content of the metadata,
the user would then request resolution of the hashed metadata 
via an explicit request at the {\em Metadata Repository}. 
In the cases of authorized modifications of the asset metadata,
the {\em Bilateral Synchronization} mechanism must
correspondingly update the NFT attribute 
with the new hash of the new asset metadata.

\item {\bf {\em Consequences}}: It must not be possible 
to retrieve asset-related information by simply reading NFT on-chain information. 
The {\em Metadata Repository} must ensure that it computes
the correct hashes from serialized metadata,
and correctly incorporates the hashes into the NFT attribute.
Thus, only these hashes are displayed on-chain.
Any user or entity seeking to access the metadata
must explicit call the {\em Metadata Repository} and supplying
it with the hash -- preferably at a protected API that requires them
to be authenticated and their access permissions verified.

Note that the utilization of these hashes in the NFT attributes
still renders the smart contract -- that manages the {\em Hidden Metadata NFT} --
to be in compliance with common NFT specifications (such as ERC721).
However,
in this pattern the platforms or marketplaces
seeking access to metadata
must be prepared to adapt their call procedures
to interact with the {\em Metadata Repository},
wielding the appropriate access credentials.

\end{itemize}

\subsection{Zero-Value NFT pattern}

\begin{itemize}[topsep=4pt,itemsep=1pt, partopsep=4pt, parsep=4pt]

\item {\bf {\em Zero-Value NFT}}: Some off-chain (and often offline) assets may follow more complex supply chain, taxation, custom clearance, or safety and security clearance procedures. In such cases the aim of an on-chain representation of the asset with NFTs is the mere proof of existence of the asset. In other words, NFTs may exist for the sole purpose of identification of the asset. {\em Product Passports}~\cite{EU-Circular-Economy-2022}
 that provide information about products in the material, construction, agri-food, seafood sectors are typical examples of such need.      

\item {\bf {\em Problem}}: A NFT with no tradeable value is needed to cover authenticity or existence of an off-chain asset. 

\item {\bf {\em Solution}}: {\em Zero-Value NFTs} are engineering construct for the purpose of maintaining consistency in the identification of assets. The smart contract managing the NFT would manage the lifecycle of the NFT by disabling any trading against payment function of the NFT. 

\item {\bf {\em Consequences}}: {\em Zero-Value NFTs} are by definition free from any economic value. Sometimes {\em Zero-Value NFTs} would also be immutable, therefore the transfer policy of the smart contract may disable transfer of ownership. In general, other characteristics of the NFT, like hiding asset metadata or making the NFT cross-chain transferable could be combined with the Zero-value nature of the NFT. 

\end{itemize}

\subsection{Discussion}
Given the rapid expansion in terms of business use cases it is clear that the above pattern list is far from being exhaustive. Other needs, like the ability of NFTs to maintain state on-chain or the ability to simultaneously manage several NFTs to cover multiple parallel facets of the lifecycle of an asset (trading, taxation, regulatory constraints, etc.) in a consistent way are just a few examples. We plan to further elaborate on the topic of Asset Proxy NFT design patterns in forthcoming publications as well as in real-life implementations.

\section{Benefits of the Asset Proxy NFT Paradigm}

There are several benefits of the {\em Asset Proxy NFT Paradigm} approach,
notably for business scenarios which require strict control over
{\em hybrid assets}:
\begin{itemize}[topsep=4pt,itemsep=1pt, partopsep=4pt, parsep=4pt]

\item	{\em Strict control over off-chain asset state with the on-chain lifecycle of the NFT}: 
The Asset Proxy NFT Paradigm provides the means for 
the state of off-chain assets to be 
perpetually synchronized with the lifecycle of their corresponding NFT(s).
A key aspect of maintaining this consistency
is the on-chain {\em Resolution} mechanism 
and the {\em Bilateral Synchronization} mechanism.

\item {\em Breaking-down the blockchain silo}: In contrast to the current effort
by many blockchain proponents to capture users onto their own ``walled garden''
blockchains (i.e. asset lock-in),
a more scalable approach is taken in our Asset Proxy NFT Paradigm
where the location information of the asset record (off-chain)
is separated from the NFT token (on-chain).
This permits the asset-instance to be moved
a different location (different {\em Asset Custody} or {\em Metadata Repository} endpoints),
without affecting the NFT token.

\item	{\em Reduced chance of identifier collisions}:
Programmatic check of the ledger(s) for prior
uses of an alphanumeric identifier string
prevents identifier collisions in the given blockchain(s).

\item	{\em Synchronization of physical objects to on-chain presence}:
The NFT carrying an alphanumeric identifier string of the off-chain asset
(denoted by this identifier) provides assurance that the string is permanently anchored on the blockchain wrapped inside the NFT data structure (attributes). This permits business logic outside the blockchain
to use this anchor to synchronize between the state of the off-chain asset with the state of the on-chain token.

\item	{\em Assist in counterfeit detection}: For product manufacturers and brand-owners, the ability to claim a namespace and issue hybrid identifiers in that namespace is critical in protecting the intellectual property associated with the products. The Asset Proxy NFT paradigm allows them to use, for example, their public-key as a namespace boundary when creating NFTs.

\item	{\em Persistence of dual off-chain and on-chain identifier}: Certain objects (e.g. physical products) may have a long life span (e.g. 50 years),
and thus, the uniqueness of their identifiers (e.g. serial numbers) 
must be guaranteed. The Asset Proxy NFT paradigm offers this kind of life span on the digital side.

\end{itemize}

\section{Conclusions}

The primary concern in this paper has been a proper asset-centric view of assets,
notably hybrid assets that consists of some
real-world valuable goods/resources (off-chain) with the NFTs (on-chain) that represent the assets on a blockchain.
We believe that supporting hybrid assets will become a core value-proposition of the Web3 vision
as the future Internet of Value.
The composition of hybrid assets introduces several new challenges,
one being the need to ensure state consistency 
between off-chain assets and the NFTs as an on-chain
representations of these assets.

In order to solve this complex problem,
we have introduced the {\em Asset Proxy NFT paradigm}
as the means to provide a framework within which
suitable technological components can be utilized to address
specific use-cases relating to hybrid assets.

As a corollary to this new paradigm,
we explore the notion of the {\em NFT design patterns}
as a means to address
the specific aspects of certain types/classes of hybrid assets.
The patterns allow the proper reasoning about the nature of the assets,
the technical components needed to support these assets,
and how the implementations can correctly integrate with other
functions (off-chain) in the ecosystem (e.g. metadata repositories and asset custody services).
The approach permits the combination of patterns to be used to address the broader challenges
related to hybrid assets in Web3.\\



\begin{thebibliography}{10}
\providecommand{\url}[1]{#1}
\csname url@samestyle\endcsname
\providecommand{\newblock}{\relax}
\providecommand{\bibinfo}[2]{#2}
\providecommand{\BIBentrySTDinterwordspacing}{\spaceskip=0pt\relax}
\providecommand{\BIBentryALTinterwordstretchfactor}{4}
\providecommand{\BIBentryALTinterwordspacing}{\spaceskip=\fontdimen2\font plus
\BIBentryALTinterwordstretchfactor\fontdimen3\font minus
  \fontdimen4\font\relax}
\providecommand{\BIBforeignlanguage}[2]{{%
\expandafter\ifx\csname l@#1\endcsname\relax
\typeout{** WARNING: IEEEtran.bst: No hyphenation pattern has been}%
\typeout{** loaded for the language `#1'. Using the pattern for}%
\typeout{** the default language instead.}%
\else
\language=\csname l@#1\endcsname
\fi
#2}}
\providecommand{\BIBdecl}{\relax}
\BIBdecl

\bibitem{Wood2018-web3}
\BIBentryALTinterwordspacing
G.~Wood, ``{Why We Need Web 3.0},'' September 2018. [Online]. Available:
  \url{https://gavofyork.medium.com/why-we-need-web-3-0-5da4f2bf95ab}
\BIBentrySTDinterwordspacing

\bibitem{Das-NFT-Security-2021}
\BIBentryALTinterwordspacing
D.~Das, P.~Bose, N.~Ruaro, C.~Kruegel, and G.~Vigna, ``{Understanding Security
  Issues in the NFT Ecosystem},'' November 2021. [Online]. Available:
  \url{https://arxiv.org/abs/2111.08893}
\BIBentrySTDinterwordspacing

\bibitem{OReilly2021-web3}
\BIBentryALTinterwordspacing
T.~{O’Reilly}, ``{Why it is too early to get excited about Web3},'' December
  2021. [Online]. Available:
  \url{https://www.oreilly.com/radar/why-its-too-early-to-get-excited-about-web3/}
\BIBentrySTDinterwordspacing

\bibitem{Moxie2022-web3}
\BIBentryALTinterwordspacing
M.~Marlinspike, ``{My first impressions of web3},'' January 2022. [Online].
  Available: \url{https://moxie.org/2022/01/07/web3-first-impressions.html}
\BIBentrySTDinterwordspacing

\bibitem{Clark88}
D.~Clark, ``{T}he {D}esign {P}hilosophy of the {DARPA} {I}nternet
  {P}rotocols,'' \emph{{ACM} {C}omputer {C}ommunication {R}eview -- {P}roc
  {SIGCOMM 88}}, vol.~18, no.~4, pp. 106--114, August 1988.

\bibitem{SmithCearley2022-Gartner-Web3}
D.~Smith, D.~Cearley, A.~Litan, and R.~Kandaswamy, ``{Web3 and the Metaverse:
  Incomplete but Complementary Visions of the Future Internet},'' Gartner,
  {R}esearch {R}eport {G00765430}, April 2022.

\bibitem{EntrikenShirley2018}
W.~Entriken, D.~Shirley, J.~Evans, and N.~Sachs, ``{ERC-721} {N}on-{F}ungible
  {T}oken {S}tandard ({EIP~721}),'' Ethereum.org, {E}thereum {I}mprovement
  {P}roposals, January 2018, available at
  {https://eips.ethereum.org/EIPS/eip-721}.

\bibitem{miller2021}
\BIBentryALTinterwordspacing
S.~Miller, ``{ART NFTs Take the Mainstream},'' December 2021. [Online].
  Available:
  \url{https://www.rimonlaw.com/blog/2021/12/29/looking-under-the-hood-diligencing-non-fungible-tokens-nft-metadata-and-smart-contracts/}
\BIBentrySTDinterwordspacing

\bibitem{arthaus2022}
\BIBentryALTinterwordspacing
{Art~Haus}, ``{On-chain NFTs and Why They’re Better},'' January 2022.
  [Online]. Available:
  \url{https://art.haus/on-chain-nfts-and-why-theyre-better/}
\BIBentrySTDinterwordspacing

\bibitem{rfc2459}
\BIBentryALTinterwordspacing
R.~Housley, W.~Ford, W.~Polk, and D.~Solo, ``Internet {X.509} public key
  infrastructure certificate and crl profile,'' January 1999, {RFC2459}.
  [Online]. Available: \url{http://tools.ietf.org/rfc/rfc2459.txt}
\BIBentrySTDinterwordspacing

\bibitem{HardjonoLipton2021-FinTech}
\BIBentryALTinterwordspacing
T.~Hardjono, A.~Lipton, and A.~Pentland, ``{A Contract Service Provider Model
  for Virtual Assets},'' \emph{{J}ournal of {FinTech}}, vol.~1, no.~2, 2021,
  available at {https://arxiv.org/pdf/2009.07413}. [Online]. Available:
  \url{https://doi.org/10.1142/S2705109921500048}
\BIBentrySTDinterwordspacing

\bibitem{Wikipedia-ISBN}
\BIBentryALTinterwordspacing
{W}ikipedia, ``{International Standard Book Number},'' 2022. [Online].
  Available:
  \url{https://en.wikipedia.org/wiki/International_Standard_Book_Number}
\BIBentrySTDinterwordspacing

\bibitem{Hardjono2019-IEEECommsMagazine}
\BIBentryALTinterwordspacing
T.~Hardjono, ``{F}ederated {A}uthorization over {A}ccess to {P}ersonal {D}ata
  for {D}ecentralized {I}dentity {M}anagement,'' \emph{{IEEE} {C}ommunications
  {S}tandards {M}agazine -- {The Dawn of the Internet Identity Layer and the
  Role of Decentralized Identity}}, vol.~3, no.~4, pp. 32--38, December 2019.
  [Online]. Available: \url{https://doi.org/10.1109/MCOMSTD.001.1900019}
\BIBentrySTDinterwordspacing

\bibitem{caip19}
\BIBentryALTinterwordspacing
{CAIP-19}, ``{Asset Type and Asset ID Specification},'' Jan 2022. [Online].
  Available:
  \url{https://github.com/ChainAgnostic/CAIPs/blob/master/CAIPs/caip-19.md}
\BIBentrySTDinterwordspacing

\bibitem{AvrilionisHardjono2021Arxiv}
\BIBentryALTinterwordspacing
D.~Avrilionis and T.~Hardjono, ``{Towards Blockchain-enabled Open Architectures
  for Scalable Digital Asset Platforms},'' October 2021. [Online]. Available:
  \url{https://arxiv.org/abs/2110.12553}
\BIBentrySTDinterwordspacing

\bibitem{GoF}
E.~Gamma, R.~Helm,, R.~E.~Johnson, J. Vlissides, ``Design patterns: Elements of reusable object-oriented software''. Reading, Mass: Addison-Wesley, 1995.
\BIBentrySTDinterwordspacing

\bibitem{Rascouet2021-PradaBlockchain}
\BIBentryALTinterwordspacing
A.~Rascouet, ``{Louis Vuitton, Cartier, Prada Push Blockchain to Ensure
  Authenticity},'' \emph{Bloomberg}, April 2021. [Online]. Available:
  \url{https://www.bloomberg.com/news/articles/2021-04-20/lvmh-cartier-prada-push-blockchain-tool-to-lure-shoppers}
\BIBentrySTDinterwordspacing

\bibitem{Irrera2020}
\BIBentryALTinterwordspacing
A.~Irrera, ``{JPMorgan in talks to merge blockchain unit Quorum with startup
  ConsenSys},'' \emph{Reuters}, February 2020. [Online]. Available:
  \url{https://www.reuters.com/article/us-jp-morgan-blockchain-exclusive/exclusive-jpmorgan-in-talks-to-merge-blockchain-unit-quorum-with-startup-consensys-sources-idUSKBN2051AW}
\BIBentrySTDinterwordspacing

\bibitem{Miller2019-Maersk}
\BIBentryALTinterwordspacing
R.~Miller, ``{IBM}-{M}aersk blockchain shipping consortium expands to include
  other major shipping companies,'' \emph{Tech Crunch}, May 2019. [Online].
  Available:
  \url{https://techcrunch.com/2019/05/28/ibm-maersk-blockchain-shipping-consortium-expands-to-include-other-major-shipping-companies/}
\BIBentrySTDinterwordspacing

\bibitem{IETF-draft-hardjono-gateways}
\BIBentryALTinterwordspacing
T.~Hardjono, M.~Hargreaves, N.~Smith, and V.~Ramakrishna, ``{An
  Interoperability Architecture for Blockchain Gateways},'' IETF,
  Internet-Draft {draft-hardjono-blockchain-interop-arch-03}, November 2021.
  [Online]. Available:
  \url{https://datatracker.ietf.org/doc/draft-hardjono-blockchain-interop-arch/}
\BIBentrySTDinterwordspacing

\bibitem{HardjonoLipton-IEEETEMS-2019}
\BIBentryALTinterwordspacing
T.~Hardjono, A.~Lipton, and A.~Pentland, ``{T}owards an {I}nteroperability
  {A}rchitecture {B}lockchain {A}utonomous {S}ystems,'' \emph{{IEEE}
  {T}ransactions on {E}ngineering {M}anagement}, vol.~67, no.~4, pp.
  1298--1309, June 2019. [Online]. Available:
  \url{doi:10.1109/TEM.2019.2920154}
\BIBentrySTDinterwordspacing

\bibitem{EU-Circular-Economy-2022}
\BIBentryALTinterwordspacing
{European Commission}, ``{EU Circular Economy Action Plan: For a cleaner and
  more competitive Europe},'' European Commission, Tech. Rep., March 2022.
  [Online]. Available:
  \url{https://ec.europa.eu/environment/pdf/circular-economy/new_circular_economy_action_plan.pdf}
\BIBentrySTDinterwordspacing

\end{thebibliography}
\end{document}